# Power-Bandwidth Efficiency and Capacity of Wireless Feedback Communication Systems

Anatoliy Platonov
Institute of Electronic Systems, Warsaw University of Technology
Warsaw, Poland
e-mail: plat@ise.pw.edu.pl

*Abstract* – **The paper is devoted to the analysis of problems appearing in optimisation and improvement of the power-bandwidth efficiency of digital communication feedback systems (FCS). There is shown that unlike digital systems, adaptive FCS with the analogue forward transmission allow full optimisation and derivation of optimal transmission-reception algorithm approaching their efficiency to the Shannon boundary. Differences between the forward channel capacity and capacity of adaptive FCS as communication unit, as well as their influence of the power-bandwidth efficiency are considered.**

## 1. INTRODUCTION

In the last years, more and more authors formulate, as the most urgent task of communications and information theory, approaching the efficiency of wireless communication systems (CS) to their theoretical limits [1-3]. However, choice of the commonly accepted criterion of CS efficiency remains open problem. The reason is existence of a number of qualitatively different but intricately related criterions of efficiency each directly connected with the complexity, cost, energy consumption and other crucial for practice characteristics of CS. Development of analytical tools enabling optimal conjugation of these criterion meets serious difficulties.

First of all, this concerns the heterogeneity of currently used criterions of the CS performance. Recently, the most widely used criterions are:
- *bandwidth efficiency* $R/F_0$ determined by the number of bits transmitted per second per one Hz of the channels band-width $[f_0 - F_0, f_0 + F_0]$, ($f_0$ - frequency of the carrier);
- *power efficiency* determined by the "energy of bit" $E^{bit} = W^{sign}T^{bit} = W^{sign}/R^{Ch}$, or "normalised" SNR $E^{bit}/N_\xi$ where $W^{sign}$ is power of the signal at the channel output, and $N_\xi$ is the spectral power density of the channel noise;
- *bit error rate* (BER) determined as a probability of erroneous transmission of a single bit.

In the previous decades, main attention in researches was concentrated on improving the bandwidth efficiency considered as it as a principal criterion due to the rapidly growing number of users under limited spectral resources [4]. Transition into GHz diapason has weakened the bandwidth limitations. In last decade, fast development of cellular nets and mass production of wireless CS for local communication radically increased the requirements to their power efficiency. For this reason, complexity, size and cost of individual communication units or the units used as the nodes of wireless nets became the criterions not less significant for the market then their power-bandwidth efficiency.

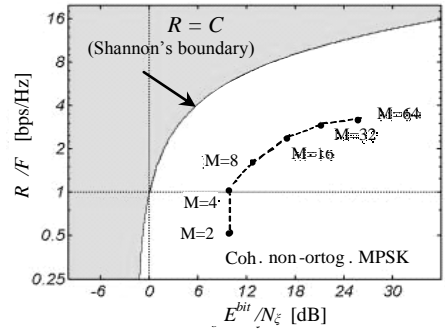

Figure 1 Illustration of the power-bandwidth tradeoff in CS with coherent non-orthogonal MPSK [1].

In turn, bit error rate criterion plays a role of additional constraint in analysis and design of CS or a subject of additional analysis supporting main researches.

One should notice that information theory quantifying upper bounds of power – bandwidth efficiency, has no developed analytical approach enabling systematic design of the systems which could work at these bounds. Nowadays, as it follows from publications (e.g. []), the most widely accepted approach to assessment of the CS performance becomes the evaluation of a degree of closeness of the points ($R/F_0$, $E^{bit}/N_\xi$) to the Shannon's boundary (Fig. 1) determined by the relationship:

$$\frac{E^{bit}}{N_\xi} = \frac{F_0}{C}\left(2^{\frac{C}{F_0}} - 1\right) \qquad (1)$$

where $C$ is the capacity of memoryless stationary channel with additive white Gaussian noise (AWGN):

$$C = F_0 \log_2\left(1 + \frac{W^{sign}}{N_\xi F_0}\right) \qquad (2)$$

Dependence (1) divides the "power-bandwidth efficiency" plain into the regions of realistic ($R < C$) and non-realisable ($R > C$) systems. Points on the line (1) correspond to the power-bandwidth efficiencies of *ideal* CS ($R = C$). In the frame of this criterions, the task of CS theory can be formulated as a development of mathematical tools permitting to design CS with the efficiencies ($E^{bit}/N_\xi$, $R/F_0$) which attain the Shannon's boundary or are close to it.

In the paper, this task is solved for the special class of CS – adaptive feedback communication systems (AFCS) with analogue signal transmission in the forward channel. The obtained results are compared with results of researches in analogue feedback CS theory carried out in years 1950-1960. We also show that the basic criterion of AFCS performance are MSE of the signal transmission and probability of appearance of rough errors which can be directly associated with BER considered in digital CS theory. Special attention is paid to the rate-distortion function. It is shown that, in optimal AFCS, this function coincides with the capacity of AFCS considered as a whole (as a specific transmission unit). New effects appearing in the work of optimal AFCS are discussed.

Solution of the task is based on the developed in [5,6] approach to optimisation of feedback estimation systems with adaptively adjusted analogue observation units. The analysis of joint transmission – receiving algorithm shows obtained as result of concurrent optimisation of the transmitting and receiving parts of AFCS its application to the system ensures its operating at the Shannon's boundary.

In the paper AFCS with the single input and single output are considered. The obtained results can be extended to multi-user CS and nets.

## 2. FULL OPTIMIZATION OF AFCS

Further, we assume that the transmitted by AFCS input signals $x_t$ are stationary Gaussian processes with known prior distribution and zero spectral power density outside the frequency interval $[-F,F]$. We assume also that each sample of the input signal is transmitted in the same way independently from previous samples. The latter one permits to reduce the analysis of system functioning to the analysis of a single sample transmission. This, in turn, permits to omit a necessity to numerate the samples in relationships presented below.

### A. AFCS as a Special Class of Estimation Systems

From the formal point of view, each AFCS (see Fig. 1) can be considered as the estimation system which delivers to addressee estimates $\hat{x}^{(m)}$ of the samples $x^{(m)}$ of input signal (further, index $m=1,2,...$ is omitted). Natural and commonly used in the applied researches criterion of estimation quality is the mean square error (MSE) of estimates [7]:

$$P_k = E[(x-\hat{x})^2] = \int_{-\infty}^{\infty}\int_{-\infty}^{\infty}[x-\hat{x}]^2 p(\hat{x}/x)p(x)\mathrm{d}\hat{x}\mathrm{d}x \quad (3)$$

where probability density function (PDF) $p(\hat{x}/x)$ can be considered as a probabilistic model of the system functioning, and $p(x)$ is the prior PDF of the samples of input signal. The form and parameters of distribution $p(\hat{x}/x)$ depend on the chosen method transmission and reception including estimates computation, as well as on the characteristics and noises of the units realizing transformation of the input signal. Optimisation of AFCS consists in definition of digital units of the transmitting and receiving parts of the system in the way minimizing MSE of transmission. Full optimisation of AFCS foresees the concurrent optimisation of both parts of the system. Unfortunately, analytical solution of this task for AFCS with digital forward transmission meets invincible mathematical difficulties. This is caused by the impossibility to express a dependence of PDF $p(\hat{x}/x)$ as the functional of possible methods of the coding and decoding, and to minimise criterion (3) over the set of coding methods. From our point of view, just this has reduced the interest to MSE as the criterion of CS performance which, as it will be shown below, is directly connected and determines the power-bandwidth efficiency of AFCS.

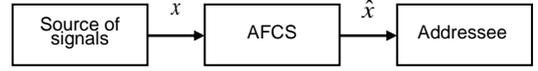

Figure 1. AFCS as the estimation system.

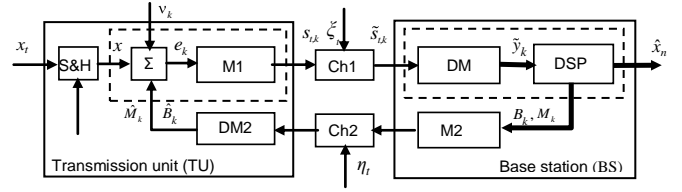

Figure 2. Block-diagram of the considered AFSC.

The results of the previous investigation [8,9] show that full optimisation of AFCS according to criterion (3) is possible in the case of adaptive analogue forward transmission. Lack of quantising and coding units enables a description of the peripheral transmission unit (TU) and the base station (BS) functioning by continuous mathematical models. This, in turn, enables analysis MSE (3) by the methods of optimal estimation theory under additional constraint on the value of permissible probability of TU overloading. This constraint enables not only adequate analysis of the system work, but also accurate analytical solution of the extreme task. In the next points, principle of solution and basic obtained results are discussed.

### B. Mathematical models

The block-diagram of considered AFCS is presented in Fig. 2. The transmitting part of TU consists of the sample and hold unit S&H, and adaptive pulse-amplitude modulator $\Sigma$+M1. We assume that the input signal $x_t$ is Gaussian with the mean value $x_0$, variance $\sigma_0^2$ and baseband [-F,F]. Each sample $x$ formed by S&H unit is transmitted iteratively in $n=T/\Delta t_0=F_0/F$ cycles ($T=1/2F$ is the sampling period, $\Delta t_0=1/2F_0$ is duration of a single cycle of transmission).

In each $k$-th cycle of transmission ($k=1,...,n$), subtractor $\Sigma$ forms the difference signal $e_k = x - \hat{B}_k$ routed to the input of modulator/emitter M1. Values $\hat{B}_k = B_k + v_k$ are estimates of the adjusting signal $B_k$ formed in BS, in digital signal processing unit (DSP) and transmitted to TU through the feedback channel Ch2. Noise $v_k$ describes the sum of feedback transmission errors and possible additive internal and external noises acting at the modulator M1 input. Below, this noise is assumed to be additive white Gaussian noise (AWGN) with the variance $\sigma_v^2$.

Like in was done [8,9], we assume that TU employs the double-side band suppressed carrier (DSB-SC) modulation (results for AFCS with SSB-AM or full AM can be obtained by recalculation of the emitted signal power). The reason of our attention to pulse-amplitude modulation (PAM) is that only this type of modulation permits direct regulation of the mean power of emitted signals. The considered model of modulator differs from the commonly used models by saturation form of transition function (Fig. 3) which enables direct consideration of possible over-modulation.

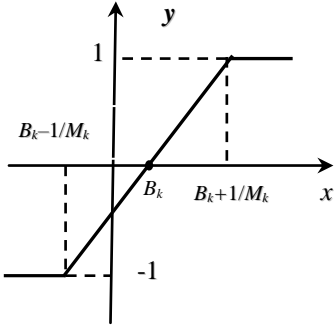

Fig. 3. Transition characteristic of adaptive modulator.

Adaptation of modulator is realised by adjusting, in each cycle, the position $B_k$ and gain $M_k$ of transition function to the values $\hat{M}_k, \hat{B}_k$ computed in DSP unit and delivered to TU through the feedback channel.
. Under these conditions, signal emitted to the forward channel Ch1 in $k$-th cycle of the sample transmission ($(k-1)\Delta t_0 \leq t \leq k\Delta t_0$) is described by the relationship:

$$s_{t,k} = A_0 \begin{Bmatrix} \hat{M}_k e_k & \text{if} & \hat{M}_k |e_k| \leq 1 \\ 1 & \text{if} & \hat{M}_k |e_k| > 1 \end{Bmatrix} \cos(2\pi f_0 t + \varphi_k) \quad (4)$$

where $A_0, f_0, \varphi_k$ are parameters of the carrier signal, and $\hat{M}_k$ are estimates of the gains $M_k$ transmitted to PTU from BS. According to [ ], in Gaussian case, optimal values of the gains $M_k$ do not depend on the signals delivered to BS in previous cycles and can be set both in TU and BS to the independently computed values. For this reason, we assume further $\hat{M}_k = M_k$.

The signal $\tilde{s}_{t,k}$ at the output of the channel Ch1 can be described by the relationship: $\tilde{s}_{t,k} = \gamma s_{t,k} + \xi_t$, where $\gamma$ is the gain coefficient of the channel, and $\xi_t$ is additive white Gaussian noise. After demodulation and digitising, signal

$$\tilde{y}_k = A \begin{Bmatrix} M_k e_k & \text{if} & M_k |e_k| \leq 1 \\ 1 & \text{if} & M_k |e_k| > 1 \end{Bmatrix} + \xi_k \quad (5)$$

is routed to the processing unit DSP. Noise $\xi_k$ in (5) is AWGN with the variance $\sigma_\xi^2 = N_\xi F_0$, where $N_\xi / 2$ is double–side spectral power density of the noise, and $A = A_0 \gamma$. Numerical coefficients in assessments of amplitude of demodulated signals, different for different types of PAM and methods of reception, can be included into $\gamma$.

In works [ ], it is shown that setting the parameters $M_k$, for each $k = 1, \ldots, n$, to the values satisfying inequality (*statistical fitting condition*):

$$\Pr_k^{over} = \Pr(M_k |e_k| > 1 \mid \tilde{y}_1^{k-1}, B_1^{k-1}, M_1^{k-1}) < \mu \quad (6)$$

determines the sets $\Omega_k$ of "permissible" values of the parameters $M_k, B_k$ which guarantee elimination of appearance of over-modulation (violation of inequality $M_k |x_k - B_k| \leq 1$) with a probability not smaller than $1 - \mu$. Values $\tilde{y}_1^{k-1} = (\tilde{y}_1, \ldots, \tilde{y}_{k-1})$ in (6) are the sequences of observations delivered to DSP of BS in previous cycles; $B_1^{k-1}, M_1^{k-1}$ are the sequences of permissible values of the modulator parameters in the previous cycles.

*Remark*: probability of the first appearance of over-modulation in $k$-th cycle has the value: $\mu + O[(k-1)\mu^2] \approx \mu$ and determines the mean percent of erroneous estimates in the sequences of estimates at the AFCS output. Assuming that undistorted estimate $\hat{x}_n$ delivers $I(X, \hat{X}_n)$ bits of information, one may consider $\mu$ as the mean percent of distorted bits or as a probability of appearance of erroneous bit (BER) in the information stream at the AFCS output.

The statistically fitted modulator works as practically linear unit, and model (4) can be replaced by the linear one that may cause the errors of the order not greater of $O(\mu)$. In this case, signals at the demodulator output can be described by the linear model:

$$\tilde{y}_k = A M_k e_k + \xi_k, \quad (e_k = x - \hat{B}_k + \nu_k) \quad (7)$$

Digital unit of BS computes, in each cycle of the sample transmission, its current estimate $\hat{x}_k = \hat{x}_k(\tilde{y}_1^k)$ according to the Kalman-type equation:

$$\hat{x}_k = \hat{x}_{k-1} + L_k[\tilde{y}_k - E(\tilde{y}_k \mid \tilde{y}_1^{k-1})] \quad (8)$$

as well as controls $B_k = B_k(\tilde{y}_1^{k-1})$ which are transmitted to TU and used for adjusting the modulator. Gains $L_k$ in (8) determine the rate of the algorithm convergence.

*C. Formulation and solution of full optimisation task*

According to Sect. 1, in each cycle of the sample transmission, we assess the performance of AFCS by MSE of estimates (3). In formulated conditions it takes the form: .

$$P_k = \int_{-\infty}^{\infty} \ldots \int_{-\infty}^{\infty} [x - \hat{x}_k(\tilde{y}_1^k)]^2 \, p(x / \tilde{y}_1^k, B_1^k, M_1^k) p(\tilde{y}_1^k) \mathrm{d}x \mathrm{d}\tilde{y}_1^k, \quad (9)$$

where $p(x / \tilde{y}_1^k, B_1^k, M_1^k)$ is the posterior probability density function of the sample values, and $\mathrm{d}\tilde{y}_1^k = \prod_{i=1}^{k} \mathrm{d}\tilde{y}_i$.

*General formulation of the optimisation task:* one should find the estimates $\hat{x}_k = \hat{x}_k(\tilde{y}_1^k)$ and the controls $B_k = B_k(\tilde{y}_1^{k-1})$, $M_k = M_k(\tilde{y}_1^{k-1})$ which, for each $k = 1, \ldots, n$, minimise MSE (9) under fulfilled fitting condition (6).

Solution of this task [8,9] has the form of algorithm determining, simultaneously, optimal rules of modulator M1 adjusting:

$$B_k = \hat{x}_{k-1} = E(x \mid \tilde{y}_1^{k-1}) \, ; \, M_k = \frac{1}{\alpha \sqrt{\sigma_\nu^2 + P_{k-1}}} \quad (10)$$

and optimal algorithm of the estimation computing :

$$\hat{x}_k = \hat{x}_{k-1} + L_k \tilde{y}_k \, ; \, L_k = M_k^{-1} \, 1 - P_k P_{k-1}^{-1} \, ; \quad (11)$$

$$P_k = P_{k-1} - \frac{A^2 M_k^2 P_{k-1}^2}{\sigma_\xi^2 + A^2 M_k^2 (\sigma_\nu^2 + P_{k-1})} =$$

$$= (1 + Q^2)^{-1} \left[ \frac{(1 + Q^2)\sigma_\nu^2 + P_{k-1}}{\sigma_\nu^2 + P_{k-1}} \right] P_{k-1}, \quad (12)$$

where

$$Q^2 = \frac{W_k^{sign}}{W^{noise}} = \frac{A_0^2 \gamma^2 M_k^2 E(e_k^2)}{\sigma_\xi^2} = \left(\frac{A}{\alpha}\right)^2 \frac{1}{N_\xi F_0} \quad (13)$$

is signal-to-noise ratio (SNR) at the analogue receiver DM1 output, and $W_k^{sign} = A / \alpha$ is the power of information component of the received signal. Saturation factor $\alpha$ is determined by the equation $\Phi(\alpha) = (1 - \mu)/2$, where $\Phi(\alpha)$ is Gaussian error function. Initial conditions: $\hat{x}_0 = x_0$; $P_0 = \sigma_0^2$.

## 3. NEW EFFECTS APPEARING IN OPTIMAL AFCS

Relationship (12) determines the low boundary of MSE of transmission (MMSE) and its analysis permits to study all main effects appearing in the work of optimal AFCS. Let us assume that SNR at the input of the modulator M1 ($SNR^{inp} = \sigma_0^2 / \sigma_v^2$) and SNR at the output of the channel Ch1 ($SNR^{Ch1} = Q^2$) satisfy the inequality

$$SNR^{inp} = \frac{\sigma_0^2}{\sigma_v^2} >> 1 + Q^2 = 1 + SNR^{Ch1} . \quad (14)$$

Under fulfilled (14), there always exists initial interval $1 < k \leq n^*$ where $P_{k-1} >> \sigma_v^2(1+Q^2) > \sigma_v^2$ and MMSE $P_k$ diminishes according to the relationship [ ]:

$$P_k = \sigma_0^2 (1+Q^2)^{-k} . \quad (15)$$

After "threshold" number of cycles $n^*$ determined by the equation $P_{k=n^*} = \sigma_v^2$, MMSE continues to diminish that results a change of the sign of inequality (14): $P_{k-1} << \sigma_v^2(1+Q^2)$. The latter reduce the rate of MMSE diminution to the hyperbolical one:

$$P_k = \frac{\sigma_v^2 P_{k-1}}{(\sigma_v^2 + P_{k-1})}\left[1+O\left(\frac{P_{k-1}}{\sigma_v^2(1+Q^2)}\right)\right] \cong \frac{\sigma_v^2 P_{k-1}}{(\sigma_v^2 + P_{k-1})} = \frac{\sigma_v^2}{n^* - k + 1} \quad (16)$$

The threshold number of cycles $n^*$ can be evaluated substituting (15) into equation $P_{n^*} = \sigma_v^2$ that results in the assessment

$$n^* = \frac{1}{\log_2(1+Q^2)} \log_2\left(\frac{\sigma_0^2}{\sigma_v^2}\right) = \frac{\log_2(SNR^{inp})}{\log_2(1+SNR^{Ch1})} . \quad (17)$$

Exponentially fast diminution of MMSE of transmission at the interval $1 < k \leq n^*$ is caused by sequential suppression of the forward channel noise due to joint optimal adjusting of modulator M1 and digital processing the received signals in digital unit of BS. This effect was noticed and studied in earlier works in analogue communications by Kailath [10], Goblick [11], Omura [12], Schalkwijk and Bluestein [13], and by many other authors. This, very intensive and promising cycle of researches, has been hampered by real successes of applied and fundamental digital communications theory which re-switched the attention of researchers. A number of works in the analogue communication theory published, in last forty years, is of a some tens order. Additional reason, which has hampered practical application of the obtained results, was linear model of modulator, not adequately describing the work of real units.

### A. Bit-rate of forward channel transmission

Using algorithm (10)-(12), one may compute corresponding entropies and the amount of information $I(\tilde{Y}_k, X_k | \tilde{Y}_1^{k-1})$ in current observation $\tilde{y}_k$ about the current value of the signal $x_k = x + v_k$ acting at the input of modulator M1 [ ]:

$$I(\tilde{Y}_k, X_k | \tilde{Y}_1^{k-1}) = H(\tilde{Y}_k | \tilde{Y}_1^{k-1}) - H(\tilde{Y}_k | X_k, \tilde{Y}_1^{k-1}) =$$
$$= \frac{1}{2}\log(1+Q^2) = \frac{1}{2}\log_2\left(1+\frac{W^{sign}}{N_\xi F_0}\right) \text{ [bit/cycle]} . \quad (18)$$

Taking into account independence of (18) from $k$, one may find the bit-rate of transmission through the channel M1-Ch1-DM1 which, under formulated conditions, takes the following form:

$$R_k^{Ch1} = \frac{I(\tilde{Y}_k, X_k | \tilde{Y}_1^{k-1})}{\Delta t_0} = F_0 \log_2\left(1+\frac{W^{sign}}{N_\xi F_0}\right) \text{ [bit/s]} . \quad (19)$$

Formula (19) coincides with general Shannon's formula for the capacity of the channel with AGWN, that is

$$R_k^{Ch1} = R_{\max}^{Ch1} = F_0 \log_2\left(1+\frac{W^{sign}}{N_\xi F_0}\right) = C . \quad (20)$$

Single difference between (20) and basic expression (2) is the way of computing the mean power of the received signals. In conventional analysis of the systems with PAM, this power is computed assuming liner models of modulators. In optimal AFCS its value is assessed as $W^{sign} = (A/\alpha)^2$ and depends on the saturation factor $\alpha$ i.e., implicitly on the accepted probability of over-modulation (BER) $\mu$.

*Corollary 1.* The obtained result shows that optimal adjusting the modulator M1 and computing the estimates of the samples according to (10) - (12) increases bit-rate in the forward channel up to its capacity independently from the feedback channel noise and the number of cycles of the sample transmission.

### B. Output bit-rate and capacity of AFCS

Below, we assume that the samples are transmitted in $n$ cycles, each during the time $n\Delta t_0 = n/2F_0$. Taking into account that the samples are transmitted independently and in the same way, and both input signals and their estimates are stationary and Gaussian, one may easily find the mean bit-rate at the AFCS output:

$$R_n^{AFCS} = \lim_{m \to \infty} \frac{I[(X^{(i)})_{i=1}^m, (\hat{X}_n^{(i)})_{i=1}^m]}{mT} = \frac{F_0}{n}\log_2\left(\frac{\sigma_0^2}{P_n}\right) \quad (21)$$

where value $I[(X^{(i)})_{i=1}^m, (\hat{X}_n^{(i)})_{i=1}^m] = mI(X, \hat{X}_n)$ is mutual amount of information in the sequence of estimates and corresponding sequence of input samples.

Substitution formulas (15), (16) into (21) gives the following relationship for optimal AFCS output bit rate:

$$R_n^{AFCS} = \begin{cases} C & \text{for } 1 \leq n \leq n^* \\ \frac{n^*}{n}\left[C + \frac{\log_2(n - n^* + 1)}{n^*}\right] & \text{for } n > n^* \end{cases} \quad (22)$$

where $C$ is the capacity of forward channel (20). Illustration of dependencies $R_n^{AFCS} = R^{AFCS}(n)$ is given in Fig. 4.

*Corollary 2.* Unlike the channel bit-rate, bit-rate at the optimal AFCS output depends on the number of cycles (on duration of the sample transmission) and has, in the "pre-threshold" interval $1 \leq n \leq n^*$, maximal values equal to the capacity of forward channel. The longer transmission ($n > n^*$) diminishes the output bit rate of the system.

The reason of this effect is exponential increase of modulation depth $M_k$ in the interval $1 \leq n \leq n^*$ what, according to (16), after $n^*$ cycles practically eliminates the influence of channel noise $\xi_k$ on MMSE $P_k$. For $n > n^*$, SNR $P_n/\sigma_v^2$ (power of informative component/feedback noise power) is much less than the unity and tends to zero for $n \to \infty$. Noise $v_k$ - sum of the feedback and other noises acting at the input of modulator becomes dominating component of the signal received by BS.

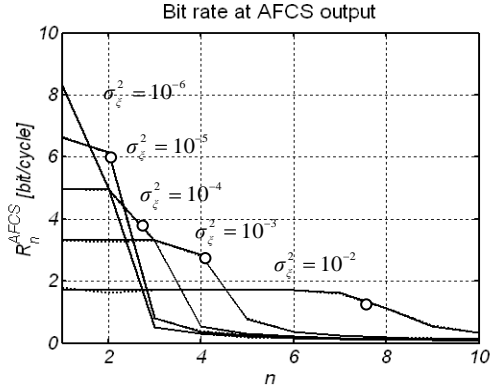

Fig. 5. Changes of output bit-rate of optimal AFCS depending on duration of the sample transmission under different powers of forward channel noise $\sigma_\xi^2$ (circles depict corresponding values of threshold number of cycles).

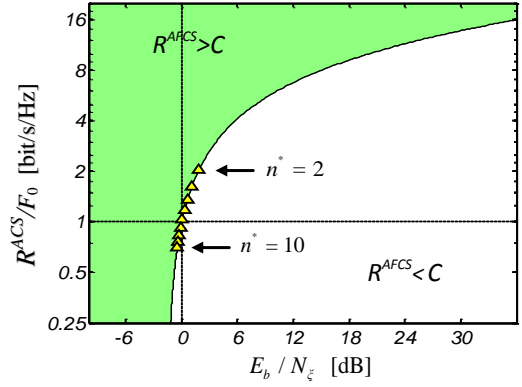

Fig. 6. Changes of power-bandwidth efficiency of optimal AFCS (triangles at the plot) depending on the threshold number of cycles [6].

Formula (21) is valid for each Gaussian signals and algorithms computing the estimates, not only for optimal ones. In turn, in optimal AFCS, $P_n$ determines low boundary of possible distortions of the signal, and formula (22) determines the upper boundary of output bit-rate. This allows us to interpret (22) as the *capacity of the system* considered as the unit. On the other side, formula (22) has the form analogous to the rate – distortion function [14] that shows direct connection of these values.

*Remark.* In AFCS with digital forward channel, quatisation of the samples and not full optimisation of the system may only decrease the values of MSE $P_n$. This allows to consider (22) as the upper boundary of bit-rates for each digital AFCS designed under same conditions as the analogue AFCS (the same transmitter power, forward channel bandwidth, signals and noises statistics).

*C. Efficiency of optimal AFCS*

Taking into account (21), one can find the energy per bit $E_n^{bit} = W^{sign}/R_n^{AFCS}$ for the sequences of estimates $\hat{x}_n$ at the AFCS output:

$$E_n^{bit} = \begin{cases} \dfrac{W^{sign}}{F_0 \log_2(1+Q^2)} = \dfrac{N_\xi Q^2}{\log_2(1+Q^2)} & \text{for } 1 \le n \le n^*; \\[2ex] \dfrac{W^{sign} n}{F_0 \left[\log_2\left(\dfrac{\sigma_0^2}{\sigma_v^2}\right) + \log_2(n-n^*+1)\right]} & \text{for } n > n^* \end{cases} \quad (23)$$

Rewriting formula (19) in the form: $Q^2 = 2^{C/F_0^{kan}} - 1$ and substituting it into the upper part of (24) permits to obtain the relationship valid for each $1 \le n \le n^*$:

$$\dfrac{E_n^{bit}}{N_\xi} = \dfrac{Q^2}{\log_2(1+Q^2)} = \dfrac{F_0}{C}\left(2^{\frac{C}{F_0}} - 1\right) \quad (25)$$

which coincides with relationship (1). For greater $n$, ($n > n^*$) value $E_n^{bit}/N_\xi$ monotonically grows, while the capacity of AFCS (22) and its bandwidth efficiency $R_n^{AFCS}/F_0$ monotonically diminish. This result allows us to formulate

*Corollary 3.* In the pre-threshold interval $[1,n^*]$, power – bandwidth efficiency of optimal AFCS realised according to algorithm (10)-(12) attains the Shannon's boundary, and systems work as the ideal communication system (see Fig. 6). For $n > n^*$, efficiency of the system monotonically decreases.

One should say that termination of transmission inside the interval $[1, n^*]$ results in MMSE $P_n$ will be much greater than $\sigma_v^2$. Although efficiency of AFCS is the same and maximal for each $1 \le n \le n^*$, resources of the system will be utilised not optimally – the same MMSE can be achieved under with much weaker requirements to the quality of feedback channel.

*Corollary 4.* The threshold point $n^*$ determines optimal number of transmission cycles ensuring full utilisation of AFCS resource.

Together with algorithm (10)-(12), the latter claim ensures full and optimal utilisation of the resources of single input-output AFCS which transmits the signals under given BER.